\documentclass{article}
\usepackage{spconf,amsmath,graphicx}
\usepackage{cite}
\usepackage{multicol,multirow}
\usepackage{booktabs}
\usepackage{colortbl}
\usepackage{arydshln}
\usepackage{caption}
\usepackage{hyperref}

\ninept
\title {Joint Pre-Training with  Speech and Bilingual Text for Direct Speech to Speech Translation}
\name{Kun Wei$^{1,\dag}$\thanks{$^{\dag}$Work done during internship at Microsoft Research Asia.}, Long Zhou$^{2}$, Ziqiang Zhang$^{2}$, Liping Chen$^2$, Shujie Liu$^2$, Lei He$^2$, Jinyu Li$^2$, Furu Wei$^2$}
\address{$^1$Audio, Speech and Language Processing Group (ASLP@NPU), School of Computer Science,\\ Northwestern Polytechnical University, Xian, China\\
     $^2$Microsoft Corporation}

\begin{document}
%
\maketitle
\begin{abstract}
Direct speech-to-speech translation (S2ST) is an attractive research topic with many advantages compared to cascaded S2ST. 
However, direct S2ST suffers from the data scarcity problem because the corpora from speech of the source language to speech of the target language are very rare.
To address this issue, we propose in this paper a Speech2S model, which is jointly pre-trained with unpaired speech and bilingual text data for direct speech-to-speech translation tasks. By effectively leveraging the paired text data, Speech2S is capable of modeling the cross-lingual speech conversion from source to target language.
We verify the performance of the proposed Speech2S on Europarl-ST and VoxPopuli datasets. Experimental results demonstrate that Speech2S gets an improvement of about 5 BLEU scores compared to encoder-only pre-training models, and achieves a competitive or even better performance than existing state-of-the-art models\footnote{Code and pre-trained models are available at \url{https://github.com/microsoft/SpeechT5/tree/main/Speech2S}.}.
\end{abstract}

\section{Introduction}
\label{sec:intro}

Direct speech to speech translation (S2ST) has gained more and more attention from research and industry communities in recent years \cite{jia2019direct,lee2021direct,popuri2022enhanced}. Traditionally, cascaded speech to speech translation consists of automatic speech recognition (ASR), machine translation (MT), and text to speech synthesis (TTS) tasks. 
Direct S2ST aims at integrating the above three tasks into an end-to-end model, which translates the speech of one language to the speech of another language directly.
Compared to cascaded S2ST, direct S2ST has the following advantages: (1) it is able to alleviate the error propagation problem of pipeline systems; (2) it can retain the emotion, pitch, and prosody information of the speaker to the greatest extent; (3) it has faster reasoning speed and takes up fewer storage resources.

However, data scarcity is the biggest problem of direct speech to speech translation tasks \cite{wang2022simple}. 
At present, there is very little parallel S2ST data  though lots of efforts \cite{wang-etal-2021-VoxPopuli,jia2022cvss, speech-matrix}.
To alleviate this problem, a line of work tries to leverage pseudo data to improve direct S2ST \cite{dong2022leveraging,popuri2022enhanced}. They usually convert the ASR data into speech to text translation data using an MT system, and then generate the target audio from the target text with a TTS system. Unfortunately, these methods do not guarantee the accuracy of the generated pseudo S2ST data. 
Another line of work aims at boosting the performance of direct S2ST through pre-training methods \cite{popuri2022enhanced,jia2022leveraging}. 
For example, the paper in \cite{jia2022leveraging} explores pre-training the encoder with mSLAM objective \cite{bapna2022mslam}, and pre-training the decoder of Translatoron 2 \cite{jia2021translatotron} with MT task to generate phonemes. The authors in \cite{popuri2022enhanced} propose to combine wav2vec 2.0 \cite{baevski2020wav2vec} encoder and mBART \cite{li2020multilingual} decoder to a speech-to-unit translation (S2UT) model, which also can be further boosted by data augmentation techniques.


Although the self-supervised pre-training method in \cite{popuri2022enhanced} can initialize the direct S2ST model with the pre-trained wav2vec 2.0 encoder and mBART decoder, which are trained with discrete units extracted with HuBERT~\cite{hsu2021HuBERT} model from unlabeled speech data,  it still lacks effective connection between encoder and decoder, and ignores the cross-lingual modeling capacity in pre-training.
In the real world, speech data, ASR data, and MT data are relative much more than direct speech to speech corpora, and MT data can be utilized to learn the transformation ability from source text to target text.
How to build the cross-lingual bridge between speech encoder and unit decoder of direct S2ST with bilingual text in the pre-training stage is not well explored.

In this paper, we propose a Speech2S model, which aims at modeling cross-lingual information and alleviating data scarcity problems by jointly pre-training with unpaired speech and bilingual MT text for the direct speech to speech translation task. 
More specially, Speech2S consists of a speech encoder, unit encoder, and unit decoder. We propose two pre-training tasks to pre-train the three modules with unit encoder as the bridge between source speech and target units. Like HuBERT \cite{hsu2021HuBERT}, the first pre-training objective is to predict the clustered units based on the output of both speech encoder and unit encoder, with unlabeled speech data. To take advantage of bilingual machine translation corpus, we first leverage two text-to-unit models to convert source/target text into source/target units, with which, the cross-lingual unit encoder and decoder can be well pre-trained through cross-entropy loss.

We evaluate the proposed model on Europarl-ST~\cite{iranzo2020europarl} and VoxPopuli \cite{wang-etal-2021-VoxPopuli} S2ST datasets.
Our contributions can be summarized as follows.
(1) We propose a joint pre-trained Speech2S model, which can take advantage of bilingual text data to boost bilingual speech conversion.
(2) The proposed model achieves a significant improvement of about 5 BLEU scores compared to the pre-trained model without MT data. 
(3) Furthermore, we conduct a detailed analysis about the effect of parallel data size, data augmentation of different domains, and subjective evaluation. 

\section{Related Work}
\label{sec:related_work}
Conventional speech to speech translation is usually composed of cascaded ASR, MT and TTS modules~\cite{nakamura2006atr,lavie1997janus}. On this basis, to avoid error transmission caused by cascade models, researchers explore the combination of ASR and MT modules~\cite{matusov2005integration,berard2016listen}, as well as TTS modules~\cite{jia2019direct,tjandra2019speech}, namely direct S2ST.
This paper focuses on exploring direct S2ST with improved pre-training methods.

\subsection{Direct Speech to Speech Translation}
S2ST, which directly translates the source speech to the target speech, has attracted a lot of attention recently~\cite{jia2019direct,tjandra2019speech,lee2021direct, kano2021Transformer,zhang2021uwspeech}. 
Translatotron \cite{jia2019direct} is the first work to achieve direct speech-to-speech translation by using a sequence-to-sequence model. This system uses an encoder to model the log-mel spectrogram and predict the target spectrogram by the decoder, combined with the speaker information. Then, a vocoder is used to convert spectrogram into waveform.
This work in \cite{jia2021translatotron} improves Translatotron system by utilzing a duration-based spectrogram synthesizer enhanced with target phoneme from decoder.
Unlike Translatotron, the authors in \cite{lee2021direct} propose a novel direct speech to speech translation system, which employs discrete hidden units instead of spectrogram as model target  before vocoder. They also expand it without using any text data on real-world S2ST tasks \cite{lee2021textless}. However, real speech to speech translation data is
very limited due to the high cost of obtaining such data \cite{wang-etal-2021-VoxPopuli,jia2022cvss}. Our work is to leverage a pre-training approach to alleviate data dependence on direct S2ST dataset.

\subsection{Pre-Training for Direct S2ST}
Recent years have witnessed a great progress on pre-training techniques for direct S2ST tasks \cite{popuri2022enhanced,jia2022leveraging}.
The work in \cite{jia2022leveraging} employs speech-text joint model from mSLAM as the encoder, to generate phoneme sequence with MT task and generate spectrogram with S2ST task.
The most related work to our paper is ~\cite{popuri2022enhanced}, which 
enhances the speech-to-unit translation (S2UT) model by a wav2vec 2.0 \cite{baevski2020wav2vec} encoder and a decoder from pre-trained unit mBART \cite{liu2020multilingual}. In this S2UT model, wav2vec 2.0 is pre-trained on unlabeled audio data, and mBART leverages reduced discrete units tokenized from unlabeled audio data to train a denoised encoder-decoder model, and finally uses the mBART decoder to initialize the S2UT decoder.
However, the simple combination of wav2vec 2.0 encoder and mBART decoder  lacks cross-language modeling capabilities, which is particularly important for translation tasks.
Motivated by this, we propose to bridge the language gap by utilizing machine translation corpus to improve model pre-training for direct speech to speech translation.

\section{The Proposed Method}
\label{sec:proposed_method}
Our goal is to leverage paired machine translation corpora to bridge the semantic gap between source speech and target speech.
In this section, we will first introduce the model architecture of Speech2S, and the details of the model pre-training and fine-tuning methods.

\subsection{Structure of Speech2S}

\begin{figure}[!t]

\centering
\includegraphics[scale=0.4]{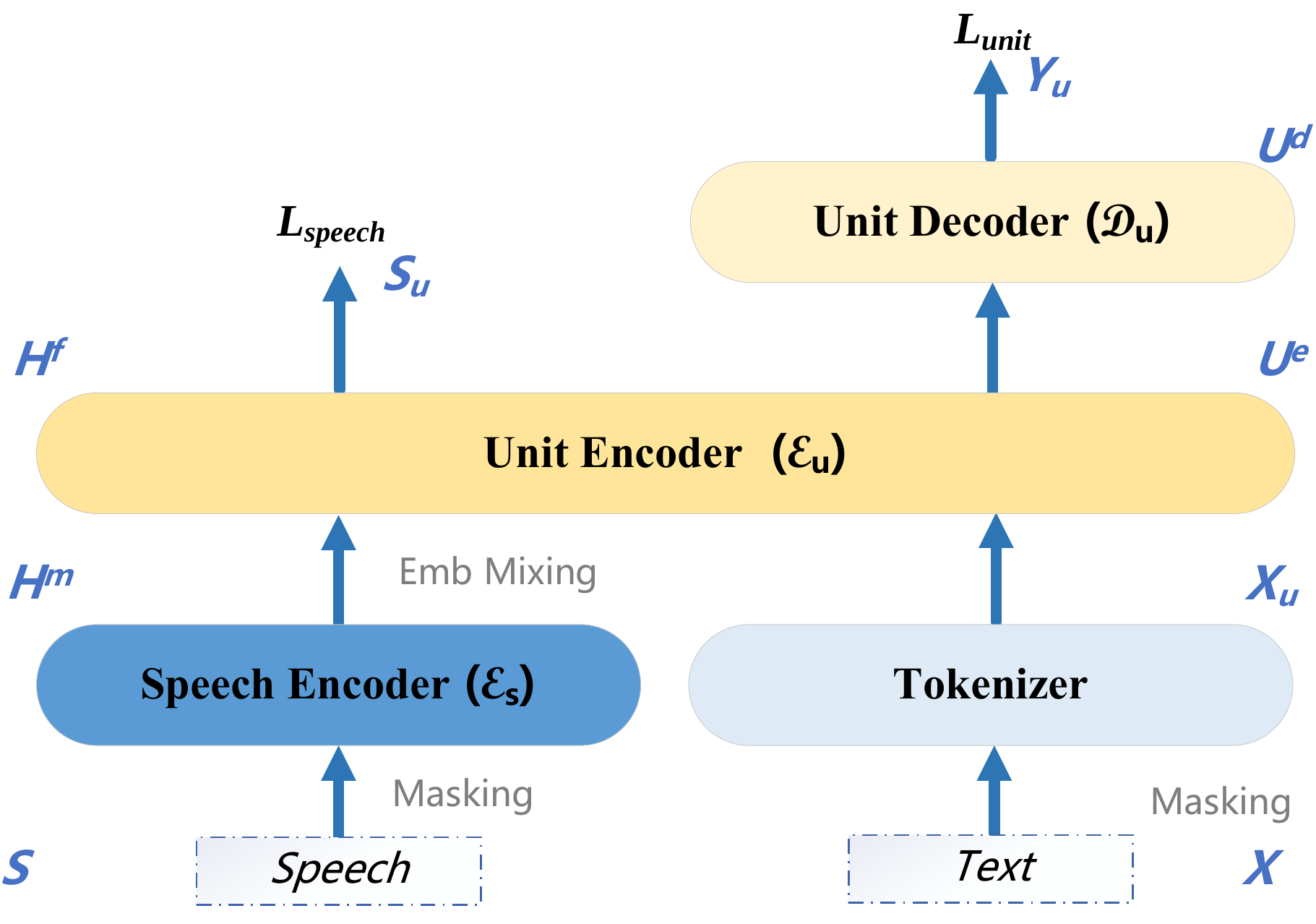}
\caption{
The overall framework of the proposed Speech2S. }
\label{fig:model}
\vspace{-0.6cm}
\end{figure}

As shown in Figure \ref{fig:model}, Speech2S consists of a speech encoder $\mathcal{E}_{s}$, a unit encoder $\mathcal{E}_{u}$ and a unit decoder $\mathcal{D}_{u}$.
Speech encoder and unit encoder employ standard Transformer network \cite{vaswani2017attention} with the same Transformer layers, except that a 5-layer CNN network in speech encoder is used to pre-process the original audio signal.
Unit decoder is a multi-layer Transformer decoder layer which is composed of a multi-head self-attention mechanism, cross-attention mechanism, and a FFN network.

Formally, we denote unpaired speech as $S$, and denote bilingual text as $(X, Y)$. After applying the speech and text discretization modules (as introduced in Section \ref{speech_text_discretization_section}), we obtain the speech units $S_u$ from $S$ and bilingual units $(X_u, Y_u)$ from $(X, Y)$. 
Briefly speaking, $\mathcal{E}_{s}$ is used to encode the source audio sequence $S$ into a sequence of vector representation $H^m$. 
Following the mixing mechanism proposed in \cite{zhang2022speechut}, we also adapt it to improve alignment learning by randomly replacing part of $H^m$ with the corresponding unit embedding.
$\mathcal{E}_{u}$ can transform speech representation $H^m$ into final hidden states $H^f$, or transform source unit sequence $X_u$ into unit hidden states $U^e$. Besides, $\mathcal{D}_{u}$ reads the encoder representations and generates a target unit sequence $Y_u$.

\subsection{Model Pre-Training}

Before pre-training, we first use two discretization modules to tokenize speech and text into shared discrete tokens.
Then the model can be optimized by two pre-training objectives, including speech to units task using speech encoder and unit encoder, and source units to target units task using unit encoder and unit decoder.

\subsubsection{Speech/text discretization}
\label{speech_text_discretization_section}

We use HuBERT k-means cluster as the speech discretization module, which is learned from the HuBERT iter-1 hidden states, and can tokenize unlabeled speech into discrete hidden units.
To tokenize text into the same space like speech, we introduce two text-to-unit models like \cite{zhang2022speechut}, which are trained by using two small ASR corpus with paired speech and transcription. More specifically, we first use speech discretization to convert paired speech into hidden units, and obtain the $\langle$text, unit$\rangle$ data by combining it with paired text. Then we utilize a sequence-to-sequence model to achieve the text-to-unit models trained on the paired text and unit data.
Once obtaining the discrete models, we can tokenize unlabeled speech $S$ into hidden units $S_u$, and tokenize bilingual text $(X, Y)$ into bilingual units $(X_u, Y_u)$, respectively, all of which can be used to optimize the model in pre-training stage.

\subsubsection{Pre-training objects}

When the input audio $S$ is fed into the speech encoder $\mathcal{E}_{s}$, it is partially masked and encoded into middle hidden states $H^m=\{h^m_1, h^m_2, ...,  h^m_T\}$, namely $\mathcal{E}_{s}$(S), which also be sent to unit encoder $\mathcal{E}_{u}$ to get final hidden states $H^f=\{h^f_1, h^f_2, ...,  h^f_T\}$ from $\mathcal{E}_{u}(H^m)$. Based on $H^m$ and $H^f$, the speech pre-training object can be designed on the masked positions as,
\begin{equation}
\begin{aligned}
  & \mathcal{L}_{speech} = - \sum_{t\in \mathcal{M}}\left(\text{log}{~p(u_t|h^m_t)} + \text{log}{~p(u_t|h^f_t)} \right)
    \label{eq:speech}
\end{aligned}
\end{equation}
where $u_t \in S_u$ is the target hidden units, and the $p(.)$ is parameterized as the same way with HuBERT \cite{hsu2021HuBERT}.

Unit encoder $\mathcal{E}_{u}$ also takes $X_u$ as input in pre-training stage, and use $\mathcal{E}_{u}(X_u)$ to output the encoded unit hidden state $U^e$. The unit decoder $\mathcal{D}_{u}$ will generate a series of hidden states $U^d$ = $\mathcal{D}_{u}(U^e)$ according to the encoder representation of source units. The objective function of unit pre-training is formalized as, 
\begin{equation}
    \mathcal{L}_{unit} = - \sum_{i=0}^{Y_u}| \text{log}~p({y_u,}_i|Y_u,_{<i}, U^e)
\end{equation}
where $y_{u,i} \in Y_{u}$, $Y_u,_{<i}$ denotes $\{y_{u,0}, y_{u,1}, ..., y_{u,i-1}\}$, and $p(.)$ is a softmax layer. Finally, we pre-train Speech2S under multi-task learning framework with $\mathcal{L}$ = $\mathcal{L}_{speech}$ + $\mathcal{L}_{unit}$.

\subsection{Speech2S Fine-Tuning}

In the fine-tuning stage, we can fine-tune Speech2S with speech encoder, unit encoder, and unit decoder to a direct speech-to-speech translation model. Leveraging the cross-entropy loss, we simply employ direct S2ST corpus as the fine-tuning dataset to optimize the model, where the target speech needs to convert into target units using speech discretization module.
Finally, we utilize a unit-based HiFi-GAN \cite{lee2021textless} to generate the target waveform from target units.

\section{Experiments}
\label{sec:experiments}
\subsection{Datasets}
We conduct our experiments on two directions of the same language pair: Spanish-English (es-en) and English-Spanish (en-es). For pre-training, we use VoxPopuli dataset, a large-scale multilingual corpus providing 100K hours of unlabelled speech data in 23 languages, as speech pre-training data. 
The ASR subset of Voxpopuli (VoxPopuli-ASR) in each language is used to train the textual discretization module, namely sequence-to-sequence based text-to-unit model. 
We use machine translation data between English and Spanish from Europarl v10~\cite{koehn2005europarl} as the bilingual text data to generate paired text units for textual unit pre-training.
Meanwhile, the speech-to-speech paired data VoxPopuli-S2S is used for our S2ST fine-tuning stage. 
We use the dev set split from VoxPopuli and the dev/test set of Europarl-ST dataset to verify the effect of speech to speech translation models. 
In order to avoid duplication with the corpus of the test set, we deleted the data of 2012 and earlier in the VoxPopuli training set. 
To avoid errors caused by audio itself, all audio is unified to the 16 kHz ogg format. In addition, we use the training sets text of CoVoST-2 and Europarl-ST datasets for additional analysis experiments on data augmentation for different domains. Data details are shown in Table 1.

\subsection{Implementation Details}
\textbf{Discretization}
We use released k-means cluster model\footnote{https://github.com/facebookresearch/fairseq/blob/main/examples/speech\\\_to\_speech/docs/textless\_s2st\_real\_data.md} from multilingual HuBERT (mHuBERT), which trained with VoxPopuli 100k subset~\cite{lee2021textless}, to extract units from speech data. 
For text discretization, we first extract the units of Voxpopuli-ASR speech using mHuBERT cluster and normalize the units using the same 1h English or Spanish speech normalizer as ~\cite{lee2021textless}. Then we train the text-to-unit discretization model using the normalized units and transcripts of the corresponding speech of the units. The text-to-unit model has 6 Transformer layers as encoder and 6 layers for decoder, each has 512 nodes with 4 attention heads.
Pairs of translation text in Europarl v10 are pre-extracted offline using this discretization model and the extracted units are applied in pre-training stage.

\textbf{Pre-training}
Our Speech2S is composed of a 6-layer Transformer speech encoder, a 6-layer Transformer unit encoder a 6-layer Transformer decoder and an output FFN layer of 1024 units.
Each Transformer layer has 768 nodes with 4 attention heads and relative positional attention bias~\cite{wang2019self}. 
We pre-train with the same 400k training steps for all models.  

\textbf{Fin-tuning}
The fine-tuning model structure is basically the same as the pre-training model structure. The normalized units of target language used in fine-tuning stage are extracted using the same extractor as the text-to-unit model.
After generating units, we use unit based HiFi-GAN~\cite{lee2021textless} to generate target speech. English and Spanish use recognition models wav2vec\footnote{https://huggingface.co/facebook/wav2vec2-large-960h-lv60-self} and microsoft speech-to-text tookit\footnote{https://azure.microsoft.com/zh-cn/products/cognitive-services/speech-to-text} to transcribe into text, respectively. The SacreBLEU toolkit~\cite{post2018call} is used to calculate the final BLEU score.

\textbf{Baselines}
For comparison, we design two strong baselines for the experiment. The first one employs HuBERT encoder to initialize the encoder of speech-to-unit translation model, and the other is existing S2UT model \cite{popuri2022enhanced}, which is initialized with HuBERT encoder plus 6-layer unit level mBART decoder. The two models use the same speech data as our model for pre-training and fine-tuning. 
The parameters of the S2UT base model and our Speech2S model are almost the same.

\begin{table}
\centering
\caption{Statistics of datasets (train/dev/test splits), including pre-training, fine-tuning, and tokenizing datasets.}
\vspace{-0.3cm}
\resizebox{\columnwidth}{!}{
\begin{tabular}{l|ccc} 
\toprule
data                                                   & samples           & source(hrs)     & target(hrs)  \\ 
\hline
\multicolumn{4}{l}{\textbf{\textit{pre-train, en-es}}}              \\ 
VoxPopuli                                              & 1.8M              & 14k             &       -       \\
Europarl v10                                           & 1.9M              &   -              &       -       \\ 
\hline
\multicolumn{4}{l}{\textbf{\textbf{\textit{pre-train, es-en}}}}         \\ 
VoxPopuli                                              & 2.0M              & 16k             &      -        \\
Europarl v10                                           & 1.9M              &   -              &      -        \\ 
\hline
\multicolumn{4}{l}{\textbf{\textit{fine-tune, en-es}}}                  \\ 
VoxPopuli-S2S                                          & 120k/6k/-         & 394/20/-        & 403/21/-     \\ 
\hline
\multicolumn{4}{l}{\textbf{\textbf{\textit{fine-tune, es-en}}}}       \\ 
VoxPopuli-S2S                                          & 153k/6k/-         & 513/19/-        & 495/18/-     \\
Europarl-ST                                            & 31.6k/1.3k/1.3k   & 75.6/3.0/2.9    & 76.5/3.0/-   \\
CoVoST-2~ ~                                            & 78.9k/13.3k/13.2k & 112.0/22.0/22.7 & 81.0/14.4/-  \\ 
\hline
\multicolumn{4}{l}{\textbf{\textit{tokenize, en}}}                \\ 
VoxPopuli-ASR                                        &          -           & 1.3k            &       -       \\ 
\hline
\multicolumn{4}{l}{\textbf{\textit{tokenize, es}}}                   \\ 
VoxPopuli-ASR                                          &     -              & 261             &     - \\       
\bottomrule
\end{tabular}
}
\vspace{-0.4cm}
\end{table}


\begin{table*}
\centering
\caption{Speech to speech translation performance (BLEU) on VoxPopuli dev set and Europarl-ST dev/test sets. For the S2UT systems, the results on VoxPopuli are reproduced by ourselves, and the results of Europarl-ST are reported in the paper.}
\vspace{-0.2cm}
\begin{tabular}{c|c|c|c|cccc} \toprule
\# & System                    & Pre-trained Model        & Parameters      & \multicolumn{2}{c}{\begin{tabular}[c]{@{}c@{}}en-es \\VoxPopuli~~Europarl-ST\end{tabular}} & \multicolumn{2}{c}{\begin{tabular}[c]{@{}c@{}}es-en \\VoxPopuli~Europarl-ST\end{tabular}}  \\ \hline
1  & \multirow{2}{*}{S2UT~\cite{popuri2022enhanced}}     & w/o pre-training & \multirow{2}{*}{Large (827M)} & -~~                    & -/21.8                                                            & -             & -/18.8                                                                     \\
2  &                           & wav2vec 2.0+mBART    &                       & ~ ~ 24.3~ ~~           & \textbf{\textbf{25.7/26.0}}                                       & ~ ~ 21.4~ ~~  & 25.7/23.8                                                                  \\ \hline
3  & \multirow{3}{*}{Ours} & HuBERT       & \multirow{3}{*}{Base (157M)} & 20.5                   & 20.2/19.1                                                         & 18.7          & 21.1/19.2                                                                  \\
4  &                           & HuBERT+mBART \cite{popuri2022enhanced} &                       & 22.5                   & 21.8/20.9                                                         & 20.1          & 23.2/21.1                                                                  \\
5  &                           & Speech2S     &                       & \textbf{\textbf{24.6}} & 25.3/25.6                                                         & \textbf{23.3} & \textbf{26.8/24.4}                                                         \\ \bottomrule
\end{tabular}
\vspace{-0.4cm}
\end{table*}








\subsection{Experimental Results}
Table 2 shows the BLEU scores of S2UT systems~\cite{popuri2022enhanced} and our Speech2S systems. By comparing the model fine-tuned from HuBERT and our proposed model, results show that our model achieves more than 4 BLEU value gains on the S2ST tasks in both directions (\#5 vs. \#3).
Compared to S2UT base model fine-tuned from HuBERT encoder and mBART decoder, the proposed Speech2S model still has an improvement of more than 3 BLEU scores  (\#5 vs. \#4). 
This result proves that our model can better incorporate text information into the language model through pre-training, and learn the corresponding relationship between source language speech and target language units through shared unit encoder. 
Furthermore, we compare our model with S2UT Large model from their paper (\#5 vs. \#2), our method achieves almost the same results as S2UT Large on the English-Spanish task with a smaller number of parameters, while on the Spanish-English test set, it achieves results that exceed those of the larger model, which also verifies the above conclusion.

\subsection{ Analysis}




\subsubsection{Effect of Parallel Data Size}
An interesting question is how well does the model perform if we only have very little fine-tuning data.
Here, we verify the effect of varying parallel data size for Speech2S and baselines.
We evaluate the proposed Speech2S and baseline from HuBERT on 10 hour, 50 hour and 100 hour supervised data sets respectively. These training data are randomly sampled from all data of VoxPopuli-S2S.

\begin{table}[h]
\vspace{-0.2cm}
\centering
\caption{BLEU scores for Speech2S and baseline trained with 15-hr, 50-hr, and 100-hr subsets.}
\vspace{-0.3cm}
\label{tab:data_size}
\begin{tabular}{c|c|cccc} 
\toprule
                       Pre-trained Model                              & hours                                            & \multicolumn{2}{c}{\begin{tabular}[c]{@{}c@{}}en-es\\dev~ test\end{tabular}}                            & \multicolumn{2}{c}{\begin{tabular}[c]{@{}c@{}}es-en\\dev~ test\end{tabular}}                             \\ 
\hline
\begin{tabular}[c]{@{}c@{}}HuBERT\\Speech2S (Ours)\end{tabular} & \begin{tabular}[c]{@{}c@{}}10\\10\end{tabular}   & \begin{tabular}[c]{@{}c@{}}0.3\\12.3\end{tabular}  & \begin{tabular}[c]{@{}c@{}}0.5\\11.9\end{tabular}  & \begin{tabular}[c]{@{}c@{}}0.5\\20.1\end{tabular}  & \begin{tabular}[c]{@{}c@{}}0.5\\19.4\end{tabular}   \\ 
\hline
\begin{tabular}[c]{@{}c@{}}HuBERT\\Speech2S (Ours)\end{tabular} & \begin{tabular}[c]{@{}c@{}}50\\50\end{tabular}   & \begin{tabular}[c]{@{}c@{}}10.2\\19.4\end{tabular} & \begin{tabular}[c]{@{}c@{}}11.2\\18.8\end{tabular} & \begin{tabular}[c]{@{}c@{}}12.6\\26.8\end{tabular} & \begin{tabular}[c]{@{}c@{}}12.9\\24.4\end{tabular}  \\ 
\hline
\begin{tabular}[c]{@{}c@{}}HuBERT\\Speech2S (Ours)\end{tabular} & \begin{tabular}[c]{@{}c@{}}100\\100\end{tabular} & \begin{tabular}[c]{@{}c@{}}12.9\\23.2\end{tabular} & \begin{tabular}[c]{@{}c@{}}13.7\\23.5\end{tabular} & \begin{tabular}[c]{@{}c@{}}15.7\\24.6\end{tabular} & \begin{tabular}[c]{@{}c@{}}14.1\\23.1\end{tabular}  \\
\bottomrule
\end{tabular}
\end{table}

From Table \ref{tab:data_size}, we can find that even if there is only 10 hours of supervised data, through our joint pre-training with speech and bilingual text, the BLEU can reach more than 10. On the 100 hour supervised data set, the fine-tuning results are close to those of hundreds of hours of supervised data fine-tuning. From the results of weak supervision, we can draw a conclusion that the Speech2S model can learn the unified mapping of speech and unit well through pre-training, thus reducing the dependence on supervised S2ST data.

\subsubsection{Effect of Data Augmentation}
In this section, we explore the effect of data augmentation for different domain datasets. As shown in Table \ref{tab:data_augmentation}, we first evaluate the performance on CoVoST-2 dev/test sets using the model trained with VoxPoluli train set. In terms of absolute performance, the BLEU scores of CoVoST-2 underperform significantly that of Europarl-ST (\#3 vs. \#1). A potential reason is that the pre-training and fine-tuning data domains are consistent for Europarl-ST test set, but it has a domain mismatch problem between VoxPopuli and CoVoST-2.


\begin{table}[h]
\centering
\caption{BLEU scores with data augmentation for different domain datasets. \textit{vp\_train} means the VoxPopuli training set, 
\textit{Eur\_train} means the Europarl-ST training set, and \textit{Cov\_train} means the CoVoST-2 training set.}
\vspace{-0.3cm}
\label{tab:data_augmentation}
\resizebox{\linewidth}{!}{%
\begin{tabular}{l|cc|cc} \toprule
\#                                           & Fine-tuning Data                                                                                       & Evaluation Data    & ~ dev~~                                                     & ~ test~~                                                     \\ \hline
\begin{tabular}[c]{@{}l@{}}1\\2\end{tabular} & \begin{tabular}[c]{@{}c@{}}~ \textit{vp\_train}~~\\~\textit{vp\_train}+\textit{Eur\_train}\end{tabular} & Europarl-ST~ & \begin{tabular}[c]{@{}c@{}}26.8\\\textbf{29.3}\end{tabular} & \begin{tabular}[c]{@{}c@{}}24.4\\\textbf{26.1}\end{tabular}  \\ \hline
\begin{tabular}[c]{@{}l@{}}3\\4\end{tabular} & \begin{tabular}[c]{@{}c@{}}~ \textit{vp\_train}\\~ \textit{vp\_train}+\textit{Cov\_train}~\end{tabular} & CoVoST-2     & \begin{tabular}[c]{@{}c@{}}15.7\\\textbf{24.2}\end{tabular} & \begin{tabular}[c]{@{}c@{}}17.6\\\textbf{26.9}\end{tabular}  \\ \bottomrule
\end{tabular}
}
\vspace{-0.4cm}
\end{table}


We conduct data augmentation experiments by adding the paired source speech and target unit data from Europral-ST and CoVoST-2 speech-to-text translation dataset. 
Based on the training data, which consists of source speech and target text, we use the text-to-unit model trained on VoxPopuli-ASR data to convert the text of the target language into units, and then enlarge the training set with the speech and generated target units, as shown in the line 2 and 4 of Table \ref{tab:data_augmentation}.
With data augmentation, the Speech2S can achieve bigger improvements on CoVoST-2 than Europarl-ST, which confirms our suspicions. Experimental results also demonstrate that this data augmentation method is very effective for domain adaption.

\subsubsection{Subjective Evaluation}
To further compare the speech quality generated by different models, we select 50 samples from the Europarl-ST dev set and test the naturalness score of these samples. 
Table \ref{tab:naturalness} lists the naturalness score of different models, including S2UT model and our Speech2S models without and with data augmentation.
The results show that our proposed Speech2S achieves the naturalness score of 4.1, outperforming S2UT model fine-tuned from HuBERT and mBART.
With data augmentation, the Speech2S model obtains the best naturalness score of 4.3.
Experiments demonstrate that our proposed method not only significantly improves the translation quality of S2ST tasks, but also enhances the naturalness of generated speech.
In addition, we can find from this experiment that more accurate units will also help to improve the quality of the final synthesized speech.

\begin{table}[h]
\centering
\caption{The naturalness score for different models. DAT means data augmentation method.}
\vspace{-0.3cm}
\label{tab:naturalness}
\begin{tabular}{c|ccc} 
\toprule
Model & S2UT   & Speech2S                         &   Speech2S+DAT                               \\ 
\hline
naturalness score & 4.0$\pm$0.1 & \begin{tabular}[c]{@{}c@{}}4.1$\pm$0.1\end{tabular} & 4.3$\pm$0.1 \begin{tabular}[c]{@{}c@{}}\end{tabular} \\ 
\hline
\end{tabular}
\vspace{-0.4cm}
\end{table}


\section{Conclusion}
\label{sec:conclusion}
This paper proposes a novel pre-training method with unlabeled speech and paired text data for direct speech to speech translation.
The core of the proposed Speech2S is to enhance the cross-lingual speech conversion capability by modeling the transformation from source units to target units, which are extracted from bilingual text data using a discrete tokenizer.
Experimental results and analyses on common VoxPopuli and Europarl-ST speech-to-speech translation tasks demonstrate the effectiveness and superiority of the proposed Speech2S model.

\small
\bibliographystyle{IEEEbib}
\bibliography{strings,refs}

\end{document}